\documentclass[aps,pra,twocolumn,showpacs]{revtex4}
\usepackage{graphicx,dcolumn}
\begin{document}
%
%
\title{ Complex extension of potentials  and  trajectories  of poles  of   the $S$-matrix element \\ in the complex momentum plane }
\author{M. Kawasaki}
\email{kawasaki@gifu-u.ac.jp}
\affiliation{Physics Department, Gifu University, Yanagido, Gifu 
501-1193, Japan}
\author{T. Maehara}
\email{tmaehar@hiroshima-u.ac.jp}
\affiliation{Graduate School  of Education, Hiroshima University, 
Higashi-Hiroshima 739-8524, Japan}
\author{M. Yonezawa}
\email{m-yonezawa@mtc.biglobe.ne.jp }
\affiliation{ Nakano 7-5-28, Aki-ku, Hiroshima 739-0321, Japan}
\date{\today}
%

\begin{abstract}
Searching for  infrastructure of the quantum mechanical system,  we study  trajectories of the $s$-wave poles of  the $S$-matrix  element   with respect to a real  phase $\alpha$ in  the  complex momentum plane  for a complex extension of   real potentials by  a   phase factor ${e}^{{i} \alpha}$.  This complex extension relates the  pole spectrum  of  the   physical system with  a potential to  the spectrum  of another system with the  potential  of  the same shape but  of  opposite sign.   There appear trajectories  with    the periodicity  of $2\pi, 4\pi$,  and  $\infty$.  The appearance  of  non-recurrent behavior of the trajectory for the change of phase  $\Delta \alpha =2\pi$ is closely related with   the existence of  resonance poles for  real repulsive potentials.  Dynamical changes of trajectory  structure are  examined.
 \end{abstract}
\pacs{03.65.Ge,03.65.Nk}
\maketitle
%
\section{Introduction}

    The hermitian requirement  for  the Hamiltonian  of  quantum mechanical system assures the reality of the energy eigenvalues with normalizable wave functions, giving the unitary system.  Extensions of hermitian  real potentials   to    non-hermitian complex  potentials   have  been made  mainly for  three  purposes.   
  
 The first   utilizes   complex potentials for  representing  phenomenologically effects of  inelastic processes on elastic scattering  known  as the optical-potential model \cite{bltt,hdgsn}. The second  is to  generalize  the  standard quantum mechanics from the standpoint that the requirement of the  reality for the  potential is too restrictive \cite{complexqm}. These two are directly associated with  non-hermitian potentials.  

The third  covers   attempts which use complex extensions as  tools of calculation for systems with \textit{real} potentials,  developed  and widely used to evaluate the  energy and width of resonances  appearing in various fields of physics \cite{msyv,riss,kkln}.   

    The infrastructure of the physical system  with complex potential has been  investigated in   its various aspects.  There seems, however,  still  to exist some structure  studied  not fully.  It will be worthwhile to clarify some properties of the solutions of Scr\"{o}dinger equation with complex potentials for further development of these studies involving complex potentials.  Here we  present an analysis  in this direction. 

We consider a system with  a  real potential   by extending it with a  phase rotation factor   $e^{i \alpha}$ where $\alpha$  is a real parameter.   We search for some global structure of the quantum system with  thus complex-extended potential.  By the phase rotation of the potential  two physical systems of  real  potentials of  opposite signatures are connected.    In other words, there is    one complex system  which interfaces  the real  world at $\alpha =n\pi$ for $n=0, \pm1, \pm2,\ldots$. 
This is  something  like  to regard  the electron state and the positron state  in the Coulomb potential as one hyper-system, though we are not intending  the extension in such a direction.  

 The pole spectrum of a system characterizes the  system.  In this paper we  examine  specifically the behavior  of poles of scattering amplitude of a particle in  potentials. By  changing  the value of  phase $\alpha$ from zero,  bound-state and antibound-state  poles  leave the imaginary axis of the complex momentum plane and  resonance poles  change their locations in the complex momentum plane.  

 Since  a  pole is  neither annihilated  nor created unless it meets other  singularities  in the complex momentum plane,   we  can unambiguously  trace   each of   poles  giving  its  trajectory.  This  establishes  the correspondence  between   the poles  generated by the  two  real potentials  which are the same except their  signatures,  exposing the pole structure for  complex potentials.

In  Sec. II we   briefly summarize some basic properties of the $S$-matrix for complex potentials.  In Sec. III   we  study stability of trajectories for  the potential strength and examine  how mutual transformations occur  between  different trajectories. Some remarks  are given in Sec. IV.

\section{Some  properties of the $S$-matrix}

We consider a  particle of mass $m$  in a complex central  scalar potential $V(r)$ with finite range.  The  time-independent  Schr\"{o}dinger equation  for the wave function  $\phi({\bf r})$ of the particle  is  in  units of $\hbar =c=1$ 
\begin{equation}
\left[-\frac{\nabla^{2}}{2m} +V(r) \right] \phi({\bf r})  = E \phi({\bf  r}) \, ,
\end{equation}
where  $E$ is the energy of the particle.

Here the complex  potential is assumed to be given by
\begin{equation}\label{eq: phase}
V(r)={e}^{{i}\alpha}V_{0}(r)\, ,
\end{equation}
where $V_{0}(r)$ is a real function of $r$  and $\alpha$ is a real phase parameter.  For $\alpha=2n\pi \,(n=0,\pm1,\pm2,...)$ the system represents the same physical  world.  Similarly,  for $\alpha = (2n+1)\pi$ we have the  system  characterized  by  the potential of the same shape and  of   opposite sign.

  We expand the wave function $\phi({\bf r})$ into the partial waves  and  denote  the  radial part of the wave function for the state with angular momentum $l$  by $u_{l}(r)/r$.  The equation for  $u_{l}(r)$ is
\begin{equation}
\biggl\{ \frac{\mathrm{d}^{2}}{\mathrm{d} r^{2}} +2m\bigl[ E -V(r) \bigr] -\frac{l(l+1)}{r^{2}} \biggr\} u_{l}(r)=0 \,.
\end{equation}

  Let $u_{l}^{(+)}(r)$ and $u_{l}^{(-)}(r)$ are  the outgoing and  the incoming wave solution, respectively.  The wave function $u_{l}(r)$  in the outer interaction-free region is given  in terms of  $u_{l}^{(+)}(r)$ and $u_{l}^{(-)}(r)$ by
\begin{equation}
u_{l}(r) = S_{l} u_{l}^{(+)}(r) - u_{l}^{(-)}(r) \,,
\end{equation}
where  $S_{l}$ is the $S$-matrix element for the angular momentum $l$.

Here we  summarize  some analytic properties of the $S$-matrices for  the  \textit{complex finite-range}  potentials \cite{kkln}.  The  momentum of the particle   $k$ outside  the range of the  potential is  given by $k=(2mE)^{1/2}$.  We study the pole structure of the $S$-matrix element in the complex momentum $k$ plane.

 For the complex finite-range  potential the $S$-matrix element with the angular momentum $l$ can be expressed by using the Jost function $f_{l}(k, \alpha)$ by
\begin{equation}
S_{l}(k, \alpha) =\frac{f_{l}(k,\alpha)}{f_{l}(-k, \alpha)}\,.
\end{equation}

 The Jost function $f_{l}(k, \alpha)$    is an entire  function of $k$ for strictly finite-range potential and has no singularities except  at infinity.  The pole of the $S$-matrix comes from the zero of $f_{l}(-k, \alpha)$. The $S$-matrix element satisfies the following  relations

\begin{eqnarray}
& & S_{l}(k, \alpha) \,\, S_{l}(-k, \alpha)  =1 \, , \label{eq: rel1} 
\\
& & S_{l}^{*}(k, -\alpha)   =  S_{l}(-k^{*}, \alpha\bmod2\pi)  \,. \label{eq: rel2} 
\end{eqnarray}
In the complex momentum plane the  pole trajectories  are  symmetric  with respect to the imaginary axis.  A pole (zero)  at the momentum $k$ implies a zero  (pole ) at $-k$. 

Although we perform the following analysis in the momentum plane, we shortly note  on the complex energy plane.  In the complex $E$ plane there is a  cut  starting  from  the branch point  $E=0$ to $\infty$.  The Riemann sheet  with  Im$\,k > 0$  is denoted by  $ I$ and  the sheet with Im$\,k < 0$  by  $II$, which are called the physical and the unphysical  sheet, respectively. 
The zeros and poles have their  complex conjugate partners on each of the physical and unphysical  sheets and  the pole trajectory is  symmetric with respect to the real  energy axis  in the complex energy plane.

 There are  two classes of poles of  the scattering amplitude;   fixed poles and  moving poles.  The  fixed-pole   spectrum depends only on the  shape of the potential and  \textit{independent} of  the potential strength as well as  the phase parameter $\alpha$, while   the location of the moving   pole   changes in the complex momentum plane with  the potential  strength  and   the phase parameter  $\alpha$. The fixed pole, though  appearing  on the positive imaginary axis of the momentum plane (or  the negative real  axis  in the  energy plane of physical sheet $I$),  is not  a physical bound state.    We are concerned with the trajectory of the moving pole here.

There appear three kinds of moving poles.  The first is  the \textit{bound-state} pole  on the positive imaginary axis  of the momentum plane (on the negative real axis   of  the  physical  sheet $I$), the second is  the \textit{antibound-state} pole on the negative imaginary axis (the negative real energy axis of the unphysical  sheet $II$), and the third  is  the \textit{resonance} pole in the lower half-plane  with positive real part  and   its  conjugate  pole with negative real part  (or   in the lower half-plane of the  sheet $II$ and   its  complex-conjugate partner  in the upper half-plane of this sheet) \cite{note1}. The  conjugate partner  is called  the \textit{antiresonance}-state  pole.

The moving poles are mutually transformed as the potential strength and the phase $\alpha$ change, with a finite number of the moving poles in the upper half-plane and  an infinite number of poles in the lower-half plane.  These three kinds of  poles should be regarded on the same physical level, though their effects to observable  physical processes are greatly different.

A naive expectation might be that any pole  will return to its starting point  after the change  of phase by $\Delta\alpha=2\pi$.     As we will see in the following,  this occurs only in  special  situations. This simple problem seems not to have been discussed so far to  the knowledge of the present authors.

\section{Pole trajectories}

We examine   potentials   for  which   solutions  of the Schr\"{o}dinger equations are  available  in   compact analytical  forms.  Such potentials  are, however, few and  the solutions are  given mostly  only for the  $s$-wave  at present.  Here we study  a set of    potentials with exponential tail and a finite-range constant potential which is called the square  potential here.

The potentials with exponential tail taken in  the  present analyses can be  covered by the expression
\begin{equation}\label{eq: genhulthenpt}
    V_{0}(r)  =-U \frac{\exp \left(-r/r_{0} \right)}{ \Bigl[1+c\exp\left(- r/r_{0}\right) \Bigr]}  \, .
\end{equation} 
which may be  called here the generalized  Hulth\'{e}n potential \cite{htpt}.  
  This constitutes a part of the more extended set  called the class of Eckart potentials \cite{nwtn}. 

Potentials of this form  including  the class of Eckart allow   us to obtain  the  analytical solution  for the $s$-wave  $S$-matrix element in terms of the hypergeometric function.
 Strictly speaking,  the potentials having exponential tail  at large distances are not finite-range, but  many of  properties  of  the scattering amplitudes  of  finite-range potentials are shared also by those of  the exponential-tail  potentials.

We give the trajectory in the complex $k$-plane. Owing to the existence of the  branch cut in the  $E$-plane, some  trajectories   look more complicated  in the $E$-plane than in the $k$-plane.

 As noted in the previous section,  all the trajectories might be thought of  having  the $2\pi$  periodicity.  In practice  we have $4\pi$ and even aperiodic open trajectories  as well as    $2\pi$ ones: this is caused by the appearance of the $s$-wave resonance poles for real  repulsive potentials.  The  $s$-wave resonance for the repulsive potential  is known, but  its physical  implication  seems  rarely to have been discussed  in publications.  

The transformation of the  real potential  $V_{0}(r)$ with the phase factor $e^{i \alpha}$ is the self-mapping of the set of all poles onto itself for $\Delta \alpha =\pm 2\pi$ and the set  is divided into invariant subsets, each of which corresponds to a closed or open trajectory in the complex $k$-plane.\cite{mapping}  If the potential depth  varies,  it  occurs  degeneracy of  two virtual states  or of  a pair of resonance and  complex virtual states.  This causes  rearrangements of trajectories.

 In the following analysis  $U$ is taken to be positive, giving  an attractive potential for $\alpha=0$ \cite{note2}.

Here  in the numerical calculations we take  the mass $m=940$ MeV and  the potential-range  parameter $r_{0}=1/140$ (MeV)$^{-1}$.   These values are typical hadronic scales, though we do not intend  to make any  application to  the  actual physical processes.

\subsection{ Potentials with exponential-tail}

First we comment on  some general features of trajectories given by the potentials with exponential tail.
To understand the characteristic features of the change of trajectories with respect to the potential strength $U$,  it is necessary to start from very small potential strength.  For vanishing  $U$  all the moving poles  approach  the fixed zeros on the negative imaginary momentum axis corresponding  to the fixed poles on the positive imaginary axis.

  This implies that each pole can be  specified by  the number $n$ specifying the zero
at  the momentum
\begin{equation}
k_{n}^\mathrm { FZ} =- i \frac{1}{2r_{0}} n \,\, \,\,\,(n=1,2,\dots)\,.
\end{equation}
 A pole associated with the  fixed zero at $k_{n}^\mathrm { FZ}$  is  designated for the  real attractive  sector of the potential  $(e^{i\alpha}=1)$ as $\mathrm{A}_{n}$, while the corresponding pole for the real repulsive sector $(e^{i\alpha}=-1)$ as  $\mathrm{R}_{n}$.

A trajectory of $2\pi$ periodicity is specified by a set of  $\mathrm{A}_{n}$ and $\mathrm{R}_{n'}$ and we express the trajectory as
\begin{equation}
 (\mathrm{R }_{n'} - \mathrm{A }_{n})   \,\,\,\, (n' \ge n )\,.
\end{equation}
 It will be  evident  that all the trajectories for  very small $U$  are  $ (\mathrm{R}_{n} - \mathrm{A}_{n}) $  with  the periodicity  2$\pi $.

 A trajectory of $4\pi$ periodicity  is  found to be characterized   by a set of four poles, $\mathrm{A}_{n}$, $\mathrm{A}_{n'}$, $\mathrm{R}_{2n-1}$, and  $\mathrm{R}_{2n}$   with $n < n'$.  Here   the pair  $\mathrm{R}_{2n-1}$ and  $\mathrm{R}_{2n}$ are no longer on the negative  imaginary momentum axis: these two are a pair of  resonance and antiresonance  poles.  Since these resonance and antiresonance poles are uniquely specified by the  leading pole $\mathrm{A}_{n}$, we need not to write explicitly the resonance-antiresonance  factors in specifying   a $4\pi$ trajectory,  which is expressed  simply   as
\begin{equation}
(\mathrm{A}_{n'} -  \mathrm{A}_{n}) \,\,\,\, (n' > n )\,.
\end{equation}

 The leading pole $ \mathrm{A}_{n}$ is either a bound-state pole or  an antibound-state pole, while the associated pole $ \mathrm{A}_{n'}$ is  an antibound state pole. As the potential strength $U$ increases, the associated  $ \mathrm{A}_{n'}$ is replaced by $ \mathrm{A}_{n''}$ with $n'' >n'$  successively by the rearrangement processes given below.

In general  the poles move counterclockwise   in the complex  momentum plane for increasing  $\alpha$ for the present choice of the phase factor (\ref{eq: phase}).

\subsection{ Exponential potential }

As a representative of the potentials with exponential tail, we mainly study the simple exponential potential, which has some  properties common to those of  the class of Eckart potentials.
 We examine  how the trajectories  change with the  potential strength $U$.

 Here  we  take    the  exponential potential given by $c=0$  for  Eq.\,(\ref{eq: genhulthenpt})
\begin{equation}
V_{0}(r) = - U \exp\left(-r /r_{0} \right)\,.
\end{equation}

The  $s$-wave solution for this potential is given  analytically \cite{exppt} as
\begin{equation}\label{eq: expsmtrx}
S_{0}(k,\alpha)= \biggl( \frac{\varphi}{2} \biggr)^{-2{i}\nu} \frac{\Gamma(1+{i}\nu) J_{{i}\nu}(\varphi)}{\Gamma(1-{i}\nu) J_{-{i}\nu}(\varphi)}\,,
\end{equation}
where  $\Gamma$ and $J$ are the gamma  and  the Bessel function  with  $\varphi \equiv 2 r_{0} (2mU{e}^{{i}\alpha})^{1/2}$ and $\nu \equiv 2 r_{0} k$.  In this case the bound-state  and antibound-state poles come from the zeros of the Bessel function in the denominator.  The poles of the gamma function in the numerator  are  fixed ones.
An interesting feature of trajectories of the exponential potential  is that  the winding behavior  of trajectories around the fixed zeros for small potential strength $U$.  As  is shown in  Appendix, the momentum of the  pole  near the fixed zero  $k_{n}^\mathrm { FZ}$ is given  for the exponential potential  by
\begin{equation}
k_{n}(\alpha) \approx k_{n}^\mathrm { FZ} + i \frac{1}{2 r_{0}}\epsilon
\end{equation}
where 
\begin{equation}
\epsilon =\frac{1}{n ! (n-1) !}  \left(\frac{\varphi}{2}\right)^{2n}\,.
\end{equation}

This implies that the  trajectory $ (\mathrm{R}_{n} - \mathrm{A}_{n})$  winds $n$-times around the fixed zero  $k_{n}^{\mathrm FZ}$  for the change of $\alpha$ from zero to $2\pi$.     It emerges, however,  only for $c=0$ in the case of  the generalized Hulth\'{e}n potential.  For $c \ne 0$, all the  trajectories wind only once for very small $U$, though the winding behavior  changes as $U$ increases.

The   trajectories of the exponential potential are not stable for the change of the potential strength $U$.  As  $U$ increases,  it happens that some of two  neighboring trajectories approach  each other and their   contact causes  restructuring between  two trajectories.  This restructuring   can be classified into two cases.  One is (1)  fusion and the other is (2)  rearrangement.

 These changes  of trajectories  will be observed   for all of  the  potentials with  exponential tail  except the Hulth\'{e}n potential.   Here we examine  the simple exponential potential.

\subsubsection{Fusion}

 This  is the process that two $2\pi$-periodic trajectories fuse into one $4\pi$-periodic trajectory,

\begin{eqnarray}
 ({\mathrm R}_{m'}-{ \mathrm A}_{m} )  &+ &  ( { \mathrm R}_{n'}-{ \mathrm A}_{n} ) \nonumber \\
& \to &  ( {\mathrm A}_{m} -  { \mathrm A}_{n}) \,.
\end{eqnarray}
Here we have  the empirical conditions $m=m'=2n$ and  $n'=2n-1$. 
\begin{figure}[t]
\begin{center}
\includegraphics[width=8cm]{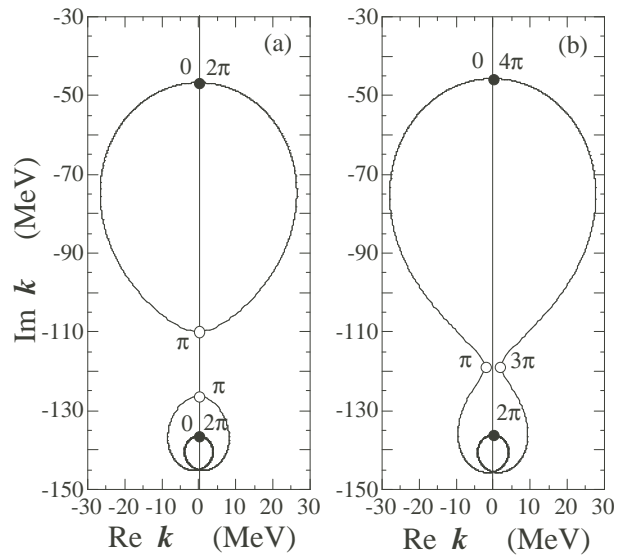}
\end{center}
\caption{An example of fusion of two  $2\pi $-periodic trajectories.  (a) Before fusion, the upper  trajectory is $(\mathrm{R}_{1}-\mathrm{A}_{1})$  and  the lower trajectory is $(\mathrm{R}_{2}-\mathrm{A}_{2})$.  Here $U$ is  4   MeV.   (b)  After fusion,  the  $4\pi$-periodic trajectory   $(\mathrm{A}_{2}-\mathrm{A}_{1})$ is produced.  Here the trajectory is shown at $U=  4.2$  MeV.  In these figures  the  closed circles  indicate the locations of  poles for the attractive sector, while the open circles for the repulsive sector. }
\label{fig: fusion1}
\end{figure}


The fusion occurs for $\phi= \pi$ on the negative imaginary momentum  axis.  An example of this process can be seen  in Fig.\,1  for $n=1$.  In Fig.\,1 (a)  the upper  trajectory is  $({ \mathrm R} _{1}-{ \mathrm A}_{1}) $ and the lower one is  $({ \mathrm R} _{2}-{ \mathrm A}_{2})$  where the winding  behavior noted above can be seen. In Fig.\,1 (b) these two  trajectories fuse into one trajectory  $( { \mathrm A} _{2}-{ \mathrm A}_{1})$.  Further increase of $U$ causes the replacement of $\mathrm{ A}_{2}$  with $\mathrm{ A}_{3}$ by a  rearrangement  processes given below.

\subsubsection{Rearrangement}
\begin{figure}[t]
\begin{center}
\includegraphics[width=8cm]{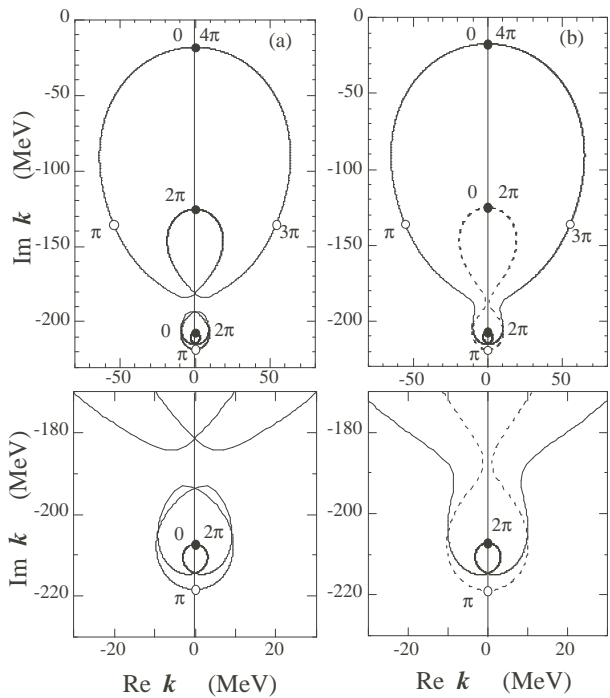}
\end{center}
\caption{ An example of rearrangement between  a $2\pi$-periodic trajectory and a  $4\pi$-periodic trajectory.  (a)  Before rearrangement, the upper trajectory is $(\mathrm{A}_{2}-\mathrm{A}_{1})$  and  the lower trajectory is $(\mathrm{R}_{3}-\mathrm{A}_{3})$.  The value of $U$ is  10.3   MeV.   (b)  After rearrangement, the upper trajectory is $(\mathrm{A}_{3}-\mathrm{A}_{1})$ (the solid curve)  and  the lower trajectory is $(\mathrm{R}_{3}-\mathrm{A}_{2})$ (the dashed curve).  The value of $U$ is  10.5   MeV.  We show below the enlarged  figures around the contact points of two trajectories.  The  closed circles  indicate the  poles for the attractive sector of the potential, and the open circles for the repulsive sector.}
\label{fig: typei1}
\end{figure}


\begin{figure}[t]
\begin{center}
\includegraphics[width=8cm]{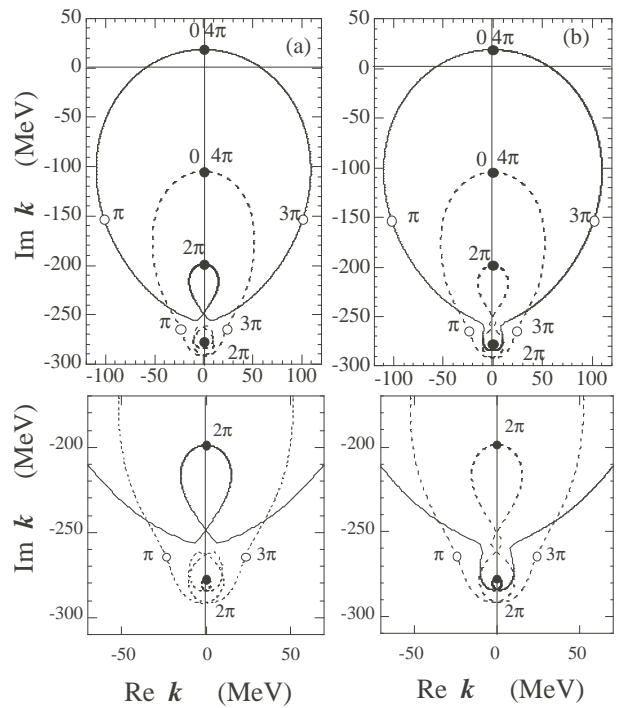}
\end{center}
\caption{An example of rearrangement between two $4\pi$-periodic trajectories. (a)  Before rearrangement,  the upper trajectory is $(\mathrm{A}_{3}-\mathrm{A}_{1})$ (the solid curve) and  the lower trajectory is $(\mathrm{A}_{4}-\mathrm{A}_{2})$ (the dashed curve). The value of $U$ is   20.4 MeV. (b) After rearrangement, the upper trajectory is $(\mathrm{A}_{4}-\mathrm{A}_{1})$  and  the lower trajectory is $(\mathrm{A}_{3}-\mathrm{A}_{2})$. The $U$ is   20.5 MeV.  Below  are shown the enlarged  parts  of the figures around the contact points of two trajectories. The  closed circles  indicate the  poles for the attractive sector of the potential, and the open circles for the repulsive sector. }
\label{fig: typeii1}
\end{figure}

There are two types of rearrangement  process, (i) and (ii).

(i) Rearrangement  between  a $2\pi$-periodic trajectory and a $4\pi$-periodic trajectory,

\begin{eqnarray}
({ \mathrm R} _{m'}- { \mathrm A}_{m} ) & + & ( { \mathrm A}_{n'} -{ \mathrm A}_{n} )  \nonumber \\
 &\to &
( { \mathrm R}_{m'}- { \mathrm A}_{n'} ) +  ( { \mathrm A}_{m} - { \mathrm A}_{n})\,
\end{eqnarray}
with   $ n < n'< m \le m' $. In the process  (i)  there occurs exchange of trajectory segments including  poles  $\mathrm{A}_{m}$ and $ \mathrm{A}_ {n'}$  between the  two trajectories.
The trajectory  $( { \mathrm A}_{n'} - { \mathrm A}_{n})$ contacts  the trajectory $( { \mathrm R}_{m'}- { \mathrm A}_{m} )$ at $\alpha = \alpha_\mathrm{c}$ with  $ \pi < \alpha_ \mathrm{c} <  2\pi  $  and  its symmetric point at   $\alpha= 4\pi - \alpha_\mathrm{c} $, while   the partner $2\pi$-periodic trajectory  $( { \mathrm R}_{m'} - { \mathrm A}_{m})$ contacts   at $\alpha_\mathrm{c} $  and  at $\alpha=2\pi -\alpha_\mathrm{c} $ owing to the  $2\pi$ periodicity of this trajectory. 
 An example of   the process (i) is given  in  Fig.\,2  for $n=1, n'=2, m=3, $ and $m'=3$.

(ii) Rearrangement  between two $4\pi$-periodic trajectories,

\begin{eqnarray}
 ({ \mathrm A}_{m'} - { \mathrm A}_{m} ) & + & ( { \mathrm A}_{n'}  - { \mathrm A}_{n} )  \nonumber \\
&\to &
( { \mathrm A}_{n'} - { \mathrm A}_{m} ) +  ( { \mathrm A}_{m'} -  { \mathrm A}_{n}) \,,
\end{eqnarray}
where  we have the conditions $ n < n',  m <  m', n < m, $ and $n' < m'$.

In the  process (ii),  the two trajectories contact  at $\alpha=\alpha_\mathrm{c'} $  with $ \pi < \alpha_\mathrm{c'} <  2\pi  $  and  its symmetric point at $\alpha=4\pi -\alpha_\mathrm{c'} $ where both trajectories develop somewhat  protruded structures  similar to those  observed  in Fig.\,2.   An example  of the process (ii) is given  in  Fig.\,3 for $n=1, n'=3, m=2,$ and $ m'=4$.

\subsection{Hulth\'{e}n potential }

The Hulth\'{e}n potential \cite{htpt} is given with $c=-1$ for Eq.\,(\ref{eq:  genhulthenpt}) as 
\begin{equation}\label{eq: hulthenpt}
V_{0}(r) = - U\frac{ \exp\left(-{r}/{r_{0}}\right)}{\Bigl[1- \exp\left(-{r}/{r_{0}}\right)\Bigr]} \,.
\end{equation}
This is the only  potential  which is singular at $r=0$ in the class of   Eckart potentials and produces  no resonance and antiresonance poles for the repulsive  sector.  The Jost function is \cite{nwtn}
\begin{equation}\label{eq: hulthenjost}
f(k,\alpha) = \frac{\Gamma(1+2{i} k r_{0} )}{\Gamma (1+{i} k r_{0} +D) \Gamma (1+{i} k r_{0} -D)}\, ,
\end{equation}
where  $D$ is defined  by $(g - k^{2} r_{0} ^{2})^{1/2}$ with  $g \equiv 2m r_{0} ^{2}Ue^{i\alpha}$. 
 If we use the infinite-product representation for  the gamma function by Euler
\begin{equation}
\Gamma (z) =  \frac{1}{z} \prod_{n=1}^{\infty} \biggl[ \bigg(1+\frac{1}{n} \bigg)^{z} \bigg(1+\frac{z}{n} \bigg) ^{-1}\biggr] \,,
\end{equation}
the Jost function (\ref{eq: hulthenjost})  is  represented as 
\begin{equation}
f(k,\alpha) =  \prod_{n=1}^{\infty} \biggl[ 1 - \frac{g}{ n(n+2{i} k r_{0} ) } \biggr] \,.
\end{equation}

The  zeros of the Jost function $f(-k,\alpha)$   give  the pole momentum    
\begin{equation} \label{eq: hulthensp}
k_{n} (\alpha)= {i}\frac{1}{2 r_{0} } \left(\frac{ 2m r_{0} ^{2}U e^{i\alpha}}{n} - n \right) \,\qquad( n=1, 2,\dots) .
\end{equation}
When $\alpha$ is zero,  the  moving poles appear  on the positive imaginary axis  for $n^{2} < g $  as  the bound states, and  on the negative imaginary axis  for  $n^{2} > g $ as the antibound states. The $S$-matrix  has   also   an infinite number of fixed  poles    coming from the poles of  the Jost function $f(k)$  on the positive imaginary axis, common to the class of Eckart potentials with exponential-tail.

Since the momentum  of antibound-state or of a  bound-state pole has an explicit expression   in this case, it is easy to trace  the pole   with respect to $\alpha$.
The trajectory is  $(\mathrm{R}_{n}-\mathrm{A}_{n})$ with $n=1,2,\dots$ and  is a circle having $2\pi$ periodicity with   its center at $ k_{n}^{\mathrm{FZ}}$ and radius $mr_{0} U/n$.
   All the pole trajectories of the Hulth\'{e}n potential are stable without  interfering with other trajectories for increase of $U$, though the  crossing   of  $(R_{n}- A_{n})$ over $(R_{n'}- A_{n'})$   with  $n<n'$  occurs.

\begin{figure}[t]
\begin{center}
\includegraphics[width=8cm]{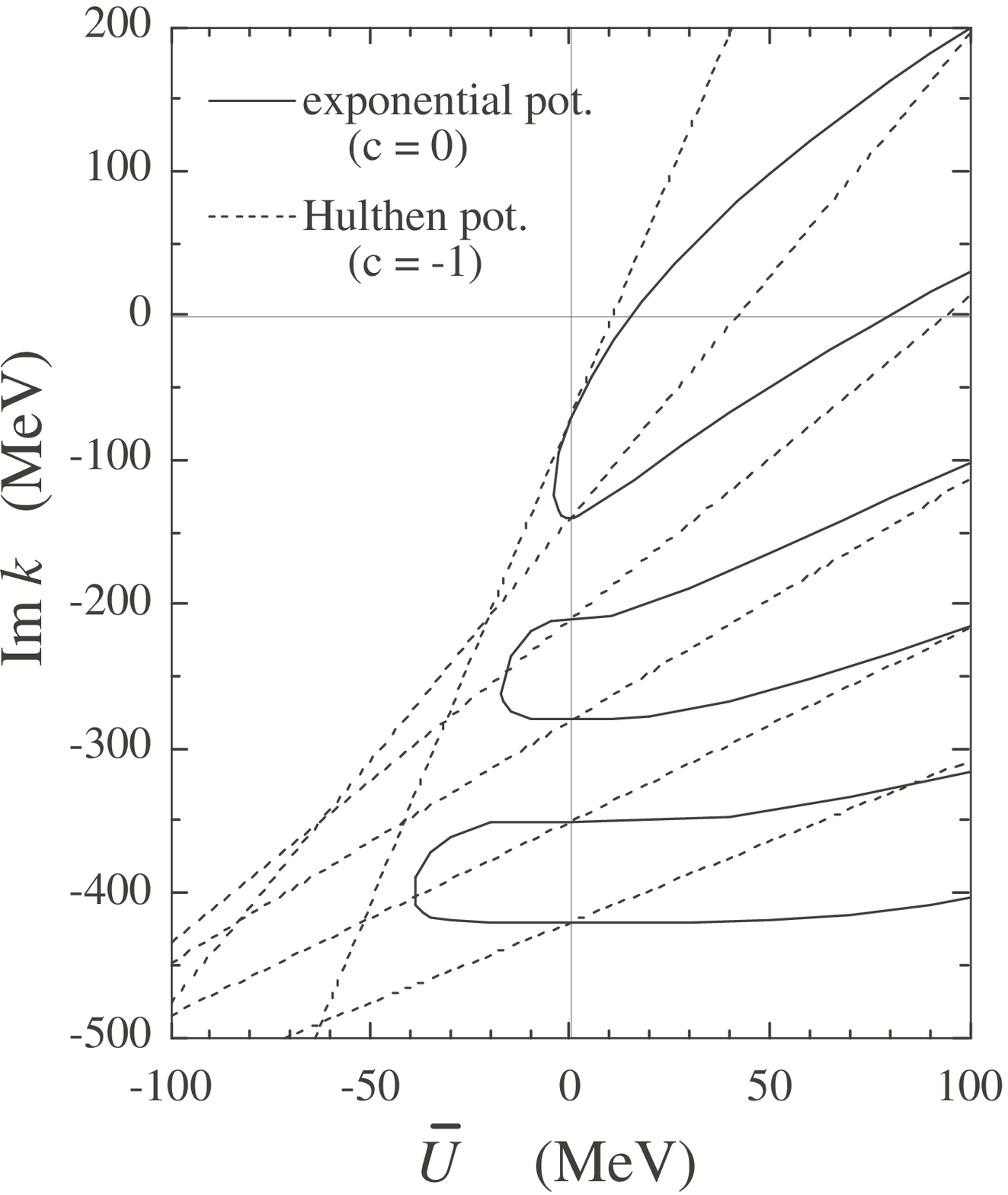}
\end{center}
\caption{The   behavior  of  moving poles   of the   exponential  $(c=0)$ and  the Hulth\'{en}   $(c=-1)$potential for   $\overline{U} \equiv  e^{i\alpha}U$ on the imaginary momentum axis.   The region   $\overline{U} >0 $ is the real attractive sector and  the region with $\overline{U} <0 $ is the repulsive one.   The  pole in the region Im\,$k >0$  is the  bound state  and  the pole  in  the region  Im\,$k<0$ is the  antibound state. }
\label{fig: A1}
\end{figure}

\begin{figure}[t]
\begin{center}
\includegraphics[width=8cm]{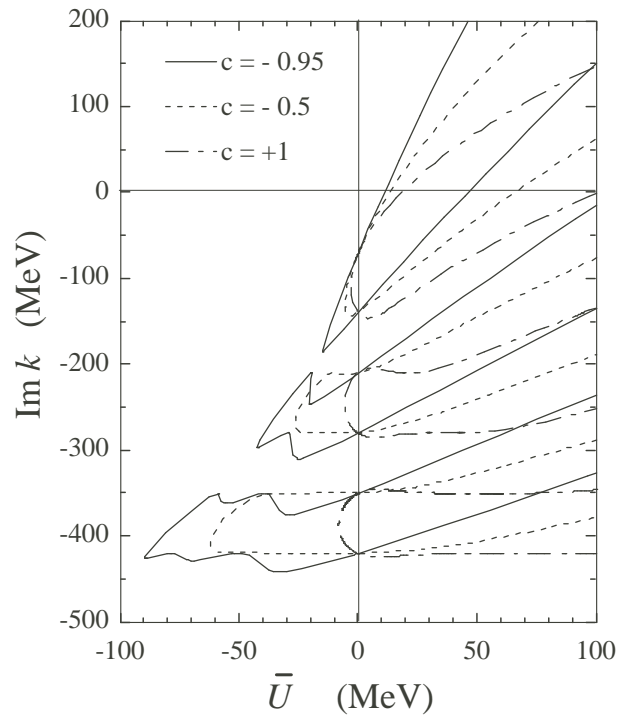}
\end{center}
\caption{The   behavior  of  moving poles   of the generalized Hulth\'{e}n potential for $c= 1.0, -0.5, -0.95$ for  $\overline{U} \equiv  e^{i\alpha}U$ on the imaginary momentum axis.   The region   $\overline{U} >0 $ is the real attractive sector and  the region with $\overline{U} <0 $ is the repulsive one.  The  pole in the region Im\,$k >0$  is the  bound state  and  the pole  in  the region  Im\,$k<0$ is the  antibound state. }
\label{fig: A2}
\end{figure}

\subsection{Formation of resonance poles by potentials  with exponential-tail }
Any trajectory  of  $4\pi$  periodicity involves a resonance  and its conjugate antiresonance poles for  the repulsive sector. 

In the weak limit of the potential strength $U$, the spectrum of antibound-state poles takes  the same  pattern both for the exponential and  the Hulth\'{e}n potentials.  If $U$  increases for the attractive sector ($e^{i\alpha}=+1$), these antibound-state poles move upward on the negative imaginary momentum axis,  and  finally appear as bound state poles on the positive imaginary axis in both  potentials. 

 For the repulsive sector ($e^{i \alpha}= -1$), the movement of poles is different between these two potentials.  In the Hulth\'{e}n  potential all the antibound-state poles move downward on the negative imaginary axis to  $- i \infty$, while in the exponential potential the top antibound-state pole moves downward  and the second moves upward, the  third  downward  and the fourth upward,  and so on.  For  $U$  continues to increase further, the top and the second collide and leave the imaginary momentum  axis, becoming  a pair of resonance  and antiresonance poles.  If   $U$  increases further,  the same happens for the third and the fourth antibound-state pole, and this continues.  The pair  of  poles R$_{n}$ and R$_{n+1} $ ($n=1,3,5,...$) become a  resonance  and its conjugate antiresonance. 

The   behavior of antibound-state poles is  shown for the exponential  and the Hulth\'{e}n potential in Fig.\,\ref{fig: A1}  about six antibound-state poles starting from $k_{n}^\mathrm{FZ}$ ($n=1- 6$) where the abscissa is the effective potential strength $\overline{U} \equiv e^{i\alpha} U $.  As $\overline{U}$ decreases, the six straight dashed lines of the Hulth\'{e}n potential  go to $-\infty$ without mutual interference, while the six  solid lines  of the exponential potential  go  to the  points of collision.

 These features   of the exponential potential are  common to  the   potentials (\ref{eq: genhulthenpt}) with exponential tail  with $c > -1$. 
The movement  of antibound-state for the repulsive sector is  monotone for the exponential, but as the potential  leaves the exponential with  $c$ different from  zero, the poles   move  in somewhat complicated ways,  but finally turn into resonance and antiresonance poles  as in  the exponential potential. 

 We show  this for some cases of the generalized Hulth\'{e}n potential  In Fig.\,(\ref{fig: A2}) where the results for $c=1,-0.5,$ and $ -0.95$ are plotted.  The collision  between two poles is always ``head-on"   with one descending  and  the other ascending  on the imaginary momentum axis.
 One might  take the curves  of $c=-0,5$  or $1$ as simple and smooth, differently from $c=-0.95$.   A detailed inspection, however,  shows  small oscillatory behavior  in these cases.

\subsection{Square potential}

The  familiar square potential gives  the pole spectrum  very different from those of exponential-tail potentials. 
We take the spherical square potential given by
\begin{equation} V_{0}(r) = \left\{ \begin{array}{rcl} 
 -U  &\qquad&  r\le r_{0} \\ 0  &\qquad& r> r_{0} \,,\end{array} \right. 
\end{equation} 
where  $U$ is a real positive constant.

Among  potentials with finite range,  the square  potential  is   the only one known  so far  for which  the Schr\"{o}dinger equation can be solved analytically    for  all   angular momentum  state at an arbitrary energy\cite{qm}.

The $S$-matrix element  for the state with the angular momentum $l$ is given by
\begin{equation}
S_{l} (k,\alpha) =  \frac{ k j_{l}(K r_{0})h_{l}^{(2)  \,\prime} (k r_{0})  - Kj_{l} ^{ \,\prime} (K r_{0})h_{l}^{(2)}(k r_{0})}{ -k j_{l}(K r_{0}) h_{l}^{(1)  \, \prime} (k r_{0})  +  K j_{l} ^{ \, \prime} (K r_{0}) h_{l}^{(1)}(k r_{0})}\,,
\end{equation}
where  $h_{l}^{(1)}$ and $h_{l}^{(2)}$ are the  spherical Hankel functions given by the spherical Bessel functions $j_{l}$ and $n_{l}$  as $h_{l}^{(1)} = j_{l} + i  \, n_{l} $ and $h_{l}^{(2)} = j_{l} - i \,  n_{l}$ with   $K \equiv [2m(E+U e^{i\alpha})]^{1/2}$.  In this case  all the poles are moving poles and there is no fixed-pole singularity. Here we examine mainly the $s$-wave and  give  a brief summary for  the  $p$-wave.  

\begin{figure}[t]
\begin{center}
\includegraphics[width=8cm]{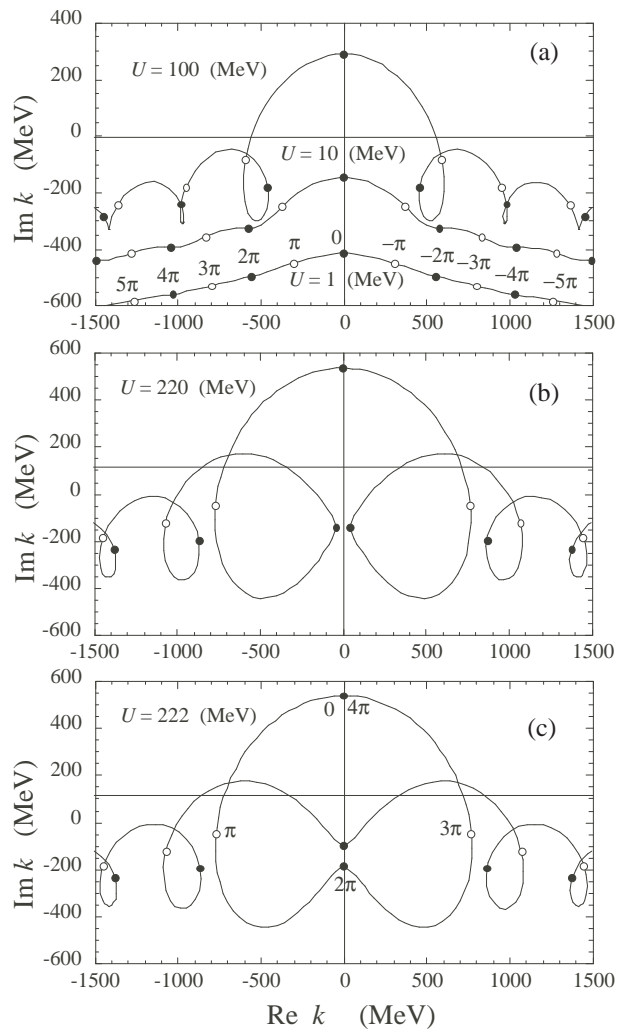}
\end{center}
\caption{The change of the trajectory of the moving pole for $U$ where  the  closed circles  indicates the locations of  poles for the attractive potential, ($\alpha =0, \pm 2\pi, ...$) while the open circles for the repulsive potential ($\alpha =\pm \pi, \pm 3\pi, ...$).  We attach the   values of $\alpha$ to the curve of $U=1$ MeV in  (a) and to those  of $U=222$ MeV  in (c). 
}
\label{fig: sqtrj}
\end{figure} 

\subsubsection{The $s$-wave trajectories  } 
The square potential produces a finite number of closed  trajectories  with $4\pi$ periodicity and  one aperiodic open trajectory.  
 For compactness, the term  \textit{attractive} is used for any pole of  the real attractive  sector of  the potential  and  \textit{repulsive} for the real  repulsive  sector.

If  the potential strength $U$ is very small,  there exists only  one  trajectory which is aperiodic  one passing  all the poles of real potentials;  both  \textit{attractive} and \textit{repulsive}  resonance and anti-resonance  poles    as well as   one  \textit{attractive} antibound-state pole  for the  change  of phase $\phi$ from $ - \infty$ to  $+ \infty$ with  $\phi=0$ at the imaginary axis.   The aperiodic open trajectory is special to the square potential. 

The critical value $U_{0, {\rm cr}}^{n}$  of  $U$  for the appearance  of the $n$th $s$-wave bound state is
\begin{equation}
U_{0,{\rm cr}}^{n} =\frac{\pi^{2}}{8 m r_{0} ^{2}}( 2n - 1)^{2} \,\,\,\,(n=1,2,\dots) \,.
\end{equation}
For $m=940$ MeV and  $ r_{0} =1/140$  (MeV)$^{-1}$ we have
\begin{equation}
U_{0,{\rm cr}} ^{n}  =25.7, \;231.5, \dots\,\,( {\rm MeV}) \,\,\,\, (n=1, 2, \dots) \,.
\end{equation}

When  $U$ increases  over  the critical value $U_{0,{\mathrm cr}}^{1}$ for the appearance of the ground bound state,  the antibound state becomes a bound state.   As  the potential strength increases further,   the  resonance pole  with $\alpha = - 2\pi$ and its  conjugate antiresonance pole  with $\alpha = + 2\pi$ come close to  and meet on the imaginary momentum axis,  turning into  two antibound-state  poles.  Then,   one of these attractive antibound-state poles  forms  a $4\pi$-periodic trajectory  with the ground bound-state pole  together with   a pair of  repulsive resonance pole  of $\alpha = - \pi$ and antiresonance pole of $\alpha = +\pi$, while the   rest of the  resonance and antiresonance poles forms  an aperiodic trajectory passing   the other attractive   antibound-state pole. 

 These changes  of the trajectories  for  $U$ are  seen in Fig.\,\ref{fig: sqtrj}.  In Fig.\,\ref{fig: sqtrj}(a)   the real  attractive and repulsive resonance and antiresonance poles  lie on a simple smooth curve with one \textit{attractive} antibound-state pole for   $U=1$ MeV, 

As $U$ increases,  all the poles move upward, approaching (parting) the imaginary axis for $\alpha = \pm 2\pi. \pm4\pi, ...  (\pm \pi, \pm 3\pi,...)$.  For  $U=100$  MeV   there is  one  bound-state pole  on the  positive imaginary axis,  which corresponds to the attractive antibound-state pole for  $U=$ 1 and    10 MeV. 

In Fig.\,\ref{fig: sqtrj}(b)  we give the trajectory for  $U=220$ MeV, just before the formation of a closed loop from the aperiodic trajectory where a pair of resonance and antiresonace poles for $\alpha = \pm 2\pi$  appear near the negative imaginary momentum axis.
In Fig.\,\ref{fig: sqtrj}(c)   we show the  two trajectories at  $U=222$ MeV,  just after the formation of a closed loop  trajectory with $4\pi$ periodicity from  the aperiodic infinite trajectory.  If  $U$ increases further,  the same process of the formation of a new closed trajectory  repeats.

\subsubsection{Notes on  the $p$-wave trajectories}

Since the square potential is the only known one for which we have the analytical solution of the $S$-matrix element  for any  angular momentum,  it will be interesting to examine the trajectory structures  of  higher partial waves.  Here we take the $p$-wave.  

The basic structures of the $p$-wave trajectories can be considered as  the same as  the $s$-wave with one infinite trajectory and a finite number of closed  trajectories for a finite value of $U$.  There are, however, some differences in details.  The formation process  of the closed trajectory is  somewhat complicated compared with the $s$-wave.

In the  $s$-wave there appears   one  \textit{attractive} antibound  pole for  very small $U$,  while there exists only  one  \textit{repulsive} antibound $p$-wave pole for an arbitrary value of $U$.  The $p$-wave trajectory is  an  aperiodic trajectory relating all the poles  including this  \textit{repulsive} antibound-state pole at $U$ smaller than the threshold value for the ground  bound-state  pole. 

 If $U$ increases,  a pair of  \textit{attractive} resonance and antiresonance poles mutually approach  and collide at  the momentum  $k=0$, then they turn into a  bound-state pole and  an antibound-state pole, which are   the ground bound-state  pole and the  ground antibound-state pole, respectively.  The ground  bound-state lies on the  aperiodic trajectories with the rest of resonance-antiresonance poles, and  the ground antibound state pole  forms a closed  loop of $2\pi$ periodicity with the  existing \textit{repulsive} antibound-state pole. 

 Further increase of $U$ causes  the contact between the aperiodic trajectory and the $2\pi$-periodic trajectory produces  a  $4\pi$-periodic trajectory passing   the ground bound-state pole and  the ground antibound-state pole  as well as  the nearest pair of repulsive resonance-antiresonace poles, while  the  repulsive antibound-state  pole is on the aperiodic trajectory of remaining resonance and antiresonance poles.     

This process of the formation of $4\pi$-periodic trajectory via the  transient $2\pi$-periodic trajectory   repeats each time  as $U$ increases over a new threshold  for excited  bound-state  pole.  There   the  collision occurs between an attractive  resonance and  its conjugate antiresonance pole at $k=0$, giving new excited bound-state and  antibound-state poles.   It  will be clear that any  $4\pi$-periodic trajectory constitutes the  bound-state and  antibound-state  poles simultaneously created by the collision.  The $4\pi$-periodic trajectories, once there are produced,  are stable for the increase of $U$ with no rearrangement with other trajectories, but the $2\pi$-periodic trajectory is an unstable intermediate product  replacing its attractive antibound-state pole each time.

For  angular momentum states higher than the $p$-wave, the basic pattern of formation of closed loops from an aperiodic trajectory  is  similar. There, however,  appear different types of  intermediate trajectories  in the  restructuring of trajectories.  For example, we meet   a pair of  $2\pi$-periodic loops which  are mutually conjugate partners and don't cross the imaginary momentum axis. Trajectories of higher waves  for   the  square potential will be discussed elsewhere.

\section{ Some Remarks}

In this paper we have searched for  some  underlying structure of the quantum system by  assuming a complex extension of  real potentials.   We have  studied  some global structure of the pole spectrum in the complex momentum plane.   It has been  found that the resonance-pole spectrum of the real repulsive potential  plays the essential role for determining the trajectory structure.   Most of potentials with exponential tail taken in this paper  produce the resonance poles for the repulsive case as well as the square potential.   The appearance of the $s$-wave resonances is not new, though seems to be discussed  very seldom.
  
 The most prevailing picture for the resonance  formation in quantum mechanics  will be  that  the resonance  is produced by  the  attractive potential in collaboration with  the centrifugal barrier. This explains the absence of the $s$-wave  resonance state  for  attractive potentials except  the square one.  Such an  intuitive interpretation is not available  for the $s $-wave resonance  which could   be purely  quantum mechanical effect, with no  classical or semi-classical  interpretations. 

The appearance of resonance poles for the real square potentials, both attractive and repulsive, can  be  attributable  to the singular nature of the potential edge.  However, even the  simple  exponential potential  produces  \textit{repulsive} resonance poles.   

 Although resonance poles appear near the real momentum axis of the physical region  for some  potentials with   resonance-type effects,  it is generally difficult to recognize the effects of $s$-wave resonances.   The poles, however, influence the observables in the physical region,  even if typical  resonance-type effects  are not produced.  It seems necessary to  exploit the  implication of the  $s$-wave resonance and  its possible role in physics. If the \textit{repulsive} $s$-wave  resonance would be  unphysical and  should be removed, this would imposes a severe restriction on the plausible type of potentials.

   In  particle quantum mechanics there is   practically  no  strong restriction for the potential.   Except a few cases, most of  potentials  assumed  in  analyses  of various processes are  of  phenomenological nature, taken for explaining  the  behavior of physical systems in some restricted area of  the energy and  the momentum transfer, and  poles remote  from the   \textit{physical} region  of interest  are  not of serious concern.  In quantum  gauge field theories,  the  interactions  among fields are  built-in  by the  gauge symmetry as well as other  invariance requirements \cite{qgft}. The study of   the quantum system with real potential   in an  extended framework as the present  one might give some clue for this  kind of problem.

All  the moving poles  have the same  dynamical origin and  we should  pay  more  attention to  the   roles of all  of the poles which characterize  the physical system.  The poles of the potentials with opposite signatures can be mutually related.  This is one reason why we study   a  hyper system with complex potential $V(r)$ with $\alpha$ changing from $-\infty$ to $\infty$.

The physical  and also  mathematical implication  of the  topological structure of trajectory shown in this article is still  to be clarified further. We hope that the present  analysis  will give a   new aspect for deeper understanding of  the  quantum  system.  

\appendix*

\section{Trajectories near the fixed zeros for the exponential potential}
As can be seen from Eq.\,(\ref{eq: expsmtrx}), the  bound-state and antibound-state poles come from the zeros of the Bessel function $J_{-i\nu}(\varphi) $.  We are interested  in how the zeros appear in the Bessel function.  In order to consider  the $n$th zero, we take the expansion of the Bessel function  in terms of  variable $ v \equiv (\varphi/2)^{2} $,
\begin{eqnarray}\label{eq: besselexp}
J_{-i \nu}(\varphi)\,e^{i\nu/2}&= & \frac{1}{\Gamma( \epsilon +1 -n)} - \frac{v}{ 1! \,\Gamma( \epsilon +2 -n )}\nonumber\\[3ex]
{}   &+& \frac{v^{2}}{2!\,\Gamma( \epsilon +3 -n )} -  \frac{ v^{3} }{3!\,\Gamma( \epsilon +4 -n )} + \cdots  \,.\nonumber\\
{} &{}&  {}
\end{eqnarray}
where we   make the substitution $ i \nu \to  n - \epsilon, \,\,\, ( |\epsilon | <1)$.

If we take up to the $n$th term in the right-hand side of Eq.\,(\ref{eq: besselexp}),  the solution of $ J_{-i \nu}(\varphi) = 0$ is given by $\epsilon = 0$ owing to the  poles of the Gamma function, irrespectively of the value of $v$. This  shows that we  cannot terminate the expansion series  at  the $n$th term.  We, therefore,   have to take at least up to  the $(n+1)$th term

\begin{figure}[t]
\begin{center}
\includegraphics[width=8cm]{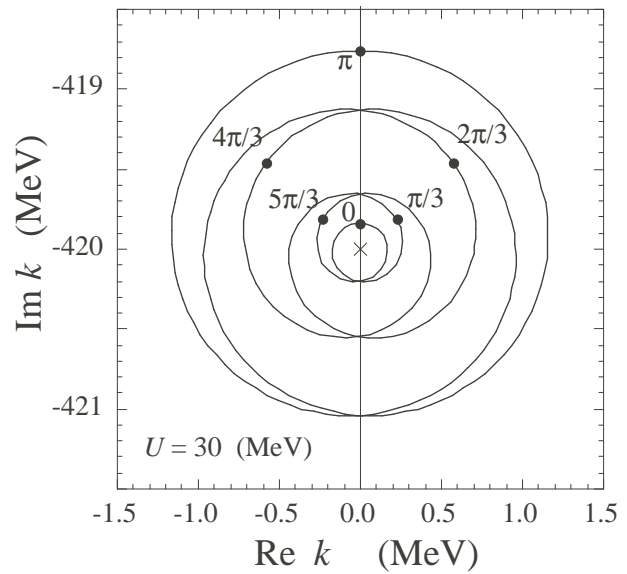}
\end{center}
\caption{An example of winding behavior of the pole trajectory around the fixed zero  for the  exponential potential.  This example  is given  for $n=6$ and $U=30$ MeV. }
\label{fig: expwd1}
\end{figure}

\begin{eqnarray}
J_{-i\nu}(\varphi)\,e^{i\nu/2} &= &\frac{1}{\Gamma( \epsilon +1 -n)}+\frac{(-v)}{ 1!\, \Gamma( \epsilon +2 -n )} + \cdots   \nonumber \\
                 & &+ \frac{ 1}{(n-1)!} \frac{(-v)^{n-1}}{\Gamma( \epsilon )} + \frac{ 1 }{n!} \frac{(-v)^{n}}{\Gamma( \epsilon +1)}  + \cdots \nonumber \\
         & \approx & \frac{1}{\Gamma( 1+\epsilon )} \Bigl\{ (\epsilon)_{n}  +  \frac{(-v)}{1!} (\epsilon)_{n-1} + \cdots   \nonumber \\
              & & + \frac{(-v)^{n-1}}{(n-1)!} (\epsilon)_{1}  +\frac{(-v)^{n}}{n!} \Bigr\} \, ,
\end{eqnarray}
where  the symbol $ (\epsilon)_{j} \equiv \epsilon (\epsilon -1) \cdots (\epsilon -j +1)  \, (j = 1,2,...)$  is used.

 The general solution of $\epsilon$ giving the zeros for the left-hand side  of this equation will be hard to  be obtained, but  here we are interested in small  $\epsilon$ as well as small $v$.   We  keep only the linear term in $\epsilon$ without $v$-dependent coefficient.  This gives the  solution for $\epsilon$ as 
\begin{equation}\label{eq: epsilon}
 \epsilon = \frac{v^{n}}{ n! (n-1)!}\,,
\end{equation}
 which vanishes  only for $v=0$.  Hence the  trajectory $ (\mathrm{R}_{n}- \mathrm{A}_{n})$  winds $n$-times around the point  $k_{n}^{\mathrm{FZ}}$  for the change of $\alpha$ from zero to $2\pi$. As  the potential strength $U$ increases,  the contributions from  higher order terms in $v$ become significant,  but  this  winding  feature remains strongly.  An example   is given   in Fig.\,7 for $n=6$ and $U=30$ MeV.

 The winding property is critically dependent on  the value of the  parameter $c$ of  the potential  (\ref{eq: genhulthenpt}) as well  as the potential strength $U$.    In general the difference of the momentum $k(\alpha)$ of  the antibound-state pole from the fixed zero, $k(\alpha) -k_{n}^\mathrm{FZ}$,  starts  linearly in $Ue^{i\alpha}$  for $c \neq 0$ and the winding number is one  for very small $U$,  while it behaves as $( Ue^{i\alpha})^{n} $ for $c=0$ .  If  $c$ is close to zero,  higher  terms in $U$  grow quickly  as $U$ increases, inducing   multiple-winding  behavior.



%
%
\end{document}